# High Harmonic Spin-Orbit Angular Momentum Generation in Crystalline Solids Preserving Multiscale Dynamical Symmetry

Kohei Nagai[1], Takuya Okamoto[1], Yasushi Shinohara[1,2], Haruki Sanada[1], and Katsuya Oguri[1]

[1] NTT Basic Research Laboratories, NTT Corporation, 3-1, Morinosato-Wakamiya, Atsugi-shi, Kanagawa 243-0198, Japan

[2] NTT Research Center for Theoretical Quantum Information, NTT Corporation, 3-1 Morinosato Wakamiya, Atsugi, Kanagawa 243-0198, Japan

e-mail: kouhei.nagai@ntt.com

**Abstract**

Symmetries essentially provide conservation rules in nonlinear light-matter interactions, that facilitate control and understanding of photon conversion processes or electron dynamics. Since anisotropic solids have rich symmetries, they are strong candidate to control both optical micro- and macroscale structures, namely spin (circular polarization) and orbital angular momentum (spiral wavefront), respectively. Here, we show structured high harmonic generation linked to the anisotropic symmetry of a solid. By strategically preserving a dynamical symmetry arising from the spin-orbit interaction of light, we generate multiple orbital angular momentum states in high-order harmonics. The experimental results exhibit the total angular momentum conservation rule of light even in the extreme nonlinear region, which is evidence that the mechanism originates from a dynamical symmetry. Our study provides a deeper understanding of multiscale nonlinear optical phenomena and a general guideline for using electronic structure to control structured light, such as through Floquet engineering.

**Introduction**

Coherent interactions between intense light fields and matter give rise to a variety of intriguing phenomena[1,2], including high harmonic generation and generation of attosecond pulses[3–5], coherent driving of electrons[6,7] and dynamical modulation of light-dressed electronic structures[8,9]. The overarching framework governing these phenomena is characterized by a spatiotemporal symmetry called dynamical symmetry (DS)[10–16]. DS serves as a powerful tool for finding universal rules in seemingly elusive and intricate phenomena arising from perturbative and non-perturbative light-matter interactions. In particular, the application of DS to microscopic light-matter interactions has provided general insights into the selection rules for polarisation of high harmonic generation (HHG)[11,13,16], as well as into symmetry breaking spectroscopy[14], molecular symmetry sensing[12,15], and light-induced symmetry-breaking phenomena[8,9,17]. For a Hamiltonian $H$ representing an electron system interacting with an external periodic light field, called a Floquet system, the DS operator $\hat{G}$ works as follows:

$$\hat{G}H(\vec{r},t)\hat{G}^{-1} = H(\hat{\gamma}_G\vec{r}, \hat{\delta}_{\hat{G}}t) = H(\vec{r},t), \quad (1)$$

where $\hat{\gamma}_G$, and $\hat{\delta}_{\hat{G}}$ are, respectively, the microscopic spatial part and temporal part of the DS operator[13]. The distinctive feature of the DS operation is that it operates not only within spatial dimensions, but also simultaneously in the temporal domain. This feature is important for describing non-perturbative phenomena, where light behaves more as a temporary oscillating electric field than individual photons.

The applicability of DS has been extended to multiscale spatial light structures, and thereby it allows for comprehensive predictions to be made on a wide range of non-perturbative optical phenomena induced by spatially non-uniform driving fields[18,19]. Micro- and macroscale light structures are respectively characterized by spin, corresponding to the helicity of the polarization, and orbital angular momentum (OAM), which corresponds to the twist in the wavefront of light[20]. This extension was motivated by the recognition of the potential significance of utilizing both of these fundamental degrees of freedom to control gas-phase HHG[21–24]. In fact, it has allowed us to develop control strategies for spin and orbital angular momentum states in extreme ultraviolet light pulses[18,19,22–26], for nonlinear beam propagation[19], and for generation of topological light[18,25,26]. However, atomic gases have only isotropic symmetry, making them poor choices for spatial design of structured light. As a result, the current strategies for designing structured light rely only on controlling the driving field to date[18,19,21–26].

Establishing a framework of DS for multiscale interactions with solid systems is of utmost importance for improving the design of structured light. Crystalline solids have anisotropic symmetry, which not only expands the design possibilities of symmetry in light-matter interactions but also has potential for using electronic structure to control in spatial structures of light, such as through Fermi-level control[27], photoinduced phase transitions[28], moiré engineering[29], and Floquet engineering[30,31]. An additional advantage is the capability of nano/ microfabrication, including metasurfaces and photonic crystals, which may lead to arbitrary control of structured harmonics[19,32,33]. One difficulty with solid systems is that, unlike isolated gas-phase atoms, they have inherently complex nonlinear interactions with light due to their dense atomic arrangements, and this complexity makes an understanding of their nature elusive. Strong light fields can induce a range of microscopic phenomena in solids including tunnelling and intraband acceleration of electrons in multiple bands, leading to nonlinear emission[34–37]. In addition, macroscopic propagation of light in bulk solids involves complex effects, such as self-phase modulation[38], cascade processes[39], and reabsorption[38]. Thus, solid systems require a predictive framework for understanding behaviours that are universal to solids, but efforts to develop such a framework have remained within the bounds of theoretical research so far[19,40].

Here, we experimentally demonstrate the generation of vectorially structured harmonics linked to the discrete crystal symmetry, starting from the concept of DS. As a platform for creating a situation characterized by multiscale DS, we made use of the spin-orbit interaction (SOI) of light in uniaxial crystals. In this situation, even from a single circularly polarized Gaussian driving beam, we observed structured harmonics composed of multiple OAM modes, and found that the conservation of total orbital angular momentum governs our observations. Our findings demonstrate that DS provides a robust framework to comprehensively explain nonlinear processes, including the complex propagation processes of nonlinear spin-orbit angular momentum cascades.

**Results**

**Multiscale dynamical symmetry and total angular momentum conservation rule of light**

Uniaxial solids can create a particular multiscale symmetry through the spin-orbit interaction (SOI) of light[41–45]. This is a striking effect because the spin and orbital angular momenta of light can be entangled in structured materials and anisotropic media[43], whereas they behave independently in isotropic media, e.g. atomic gases, for paraxial beams[20]. When a circularly polarized beam is focused onto a thick uniaxial crystal along its optical axis (Fig. 1a) [42,44], the light component traveling obliquely at an angle $\theta$ with respect to the optical axis experiences birefringence due to the anisotropic refractive indices, $n_o$ and $n_e(\theta)$, as shown in Fig. 1b. As a result, the polarization state of the light oscillates between right and left circular polarized depending on $\theta$ with its axisymmetric spatial distribution around the optical axis.

We show that multiscale DS predicts a total angular momentum conservation rule for both perturbative and nonperturbative HHG when a strong laser illuminates a uniaxial crystal under the tight focus conditions. When $\hat{G}$ is a DS operator of an electron system interacting with an external light field, the electric fields of the high harmonics emitted by the system remain identical under the same DS operation $\hat{G}$[19]. In situations where the crystal structure of the solid possesses $n$-fold rotational symmetry within the plane perpendicular to the optical axis and the laser beam has a spin and orbital angular momentum state of $(s_1, l_1)$, the system in Fig. 1a is expected to have the following two DS operators:

$$\hat{G}_1 = \hat{R}_{n,1} \hat{r}_{n,1} \hat{\tau}_{n,-s_1-l_1} \tag{2}$$

and

$$\hat{G}_2 = \hat{R}_{2,1} \hat{\tau}_{2,-l_1}. \tag{3}$$

The operator $\hat{G}_1$ is composed of three operations: a temporal translation $\hat{\tau}_{n,-s_1-l_1}$ of $-(s_1 + l_1)/n$ times the period of the light field, a microscopic spatial rotation $\hat{r}_{n,1}$ of $2\pi/n$, associated with the crystal symmetry and polarization of light, and a macroscopic spatial rotation $\hat{R}_{n,1}$ of $2\pi/n$. Here, we assume that the electric field is periodic in time and has a negligible $z$-component. Figure 1c shows an example of the spatial distribution of polarizations in a plane parallel to the x-y plane when a circularly polarized Gaussian beam is applied to a uniaxial crystal. In the previous research on solids, only microscopic DS operations (e.g. $\hat{r}$ and $\hat{\tau}$) were considered in an effort to understand nonlinear responses because a driving field with a spatially uniform circular polarization was applied[16,46]. However, the distribution of polarizations becomes non-uniform in a system with SOI, which breaks local microscopic DS. Even under such a condition, by subsequently applying the macroscopic operation of $\hat{R}$, the system becomes identical to the original. The multiscale operation $\hat{G}_1$ predicts that the total angular momentum is conserved as

$$J_m = mJ_1 + nQ, \tag{4}$$

where $J_m$ is the total angular momentum of the $m$-th order harmonics defined by $J_m = l_m + s_m$, $l_m$ and $s_m = \pm 1$

represent the indices of orbital and spin angular momenta of the $m$-th harmonics, respectively, and $Q$ is an integer. In addition, $\hat{G}_2$ predicts the following restriction:

$$l_m = ml_1 + 2Q', \qquad (5)$$

where $Q'$ is an integer. This equation restricts $l_m$ to even integers when $l_1$ takes an even integer value (see the Supplementary Information for a detailed derivation). If the laser beam interacts with gas media with spherical symmetry, the right-hand side of equation (4) becomes zero[11,23,47] for conserving the angular momenta between the incident and emitted photons. In crystalline solids, on the other hand, an $nQ$ degree of freedom arises from their discrete rotational symmetry[46,48,49]. Previous research has shown that in the absence of the SOI, the spin and orbital angular momenta are independently conserved in solids (Fig. 1d) [46,50]. On the other hand, in the presence of the SOI, it is possible to control both spin and orbital angular momentum with the combined symmetry of light and crystal.

To investigate the above processes in a real material, we spectrally and spatially measured the higher harmonics generated by focusing a laser beam tightly on a uniaxial crystal (Fig. 1e). The driver laser was a right-circularly polarized (RCP, $s_1 = 1$) infrared pulse with a photon energy of 0.51 eV and had a Gaussian-like beam profile with $l_1 = 0$ (inset of Fig. 1e). The sample was GaSe with a bandgap energy of 2.2 eV and in-plane threefold rotational symmetry ($n = 3$ in equation (4)). The GaSe crystal is a standard platform for investigating HHG[35–37] and has ideal properties for studying the SOI of light because of its significant uniaxial anisotropy ($n_e$= 2.41, $n_o$= 2.74 at 0.51 eV)[51]. Here, we strategically used a crystal with a thickness of 2 mm to induce a strong SOI. Thick crystals are not usually chosen for HHG research because of the difficulty in handling the complicated cascade processes and phase matching[38]. In all of the experiments reported below, we optimized the focus point so that the above-bandgap harmonics from the back surface of the crystal were maximized.

**Effect of spin-orbit interaction on harmonic spectra.**

The circular-polarization resolved harmonic spectra are shown in Fig. 2a. The fundamental beam was tightly focused with an external Gaussian divergence angle of 402 mrad. Both polarization components appeared in the harmonics up to sixth order. Both even and odd-order harmonics appeared because the GaSe crystal lacks inversion symmetry. Since the bandgap energy is around the photon energy of the fourth-order harmonics, we expected the conversion process to be significantly different for each order: i.e., we expected that the third order and lower harmonics would mainly be generated as the light propagated in the crystal, while the fourth order and higher harmonics would mainly be generated from the back surface due to reabsorption in the crystal. To confirm the effect of SOI, we compared the measured spectra with those of a loose focus condition with a 12.7 mrad external divergence angle, as shown in the inset of Fig. 2a. In contrast to the tight focus condition, only the RCP fourth-order harmonics and left-circularly polarized (LCP) fifth-order harmonics appeared, while the other harmonics were largely suppressed in accordance with the spin angular momentum conservation rules[46]. Thus, our observations clearly demonstrate the effect of the tight focus in mixing different polarization components. In fact, as calculated in Fig. 2b, the divergence angle of 402 mrad is large enough to generate almost equal amounts of counter-rotating circular polarization components for the fundamental beam. To confirm the nonlinearity of the observed HHG process, we measured the dependence of the harmonic intensity on the incident pulse energy (Fig. 2c). The focused intensity of the infrared driving pulse was estimated to be 0.15 TW cm$^{-2}$ at the pulse energy of 0.1 uJ cm$^{-2}$, which is enough to reach the extreme nonlinear regime[35–37]. Actually, all orders of harmonics, especially the higher order ones, deviated from the power law predicted by perturbation theory. All experiments reported below were performed in the extreme nonlinear regime (0.1 uJ cm$^{-2}$; grey line in Fig. 2c).

**Spin-dependent structured high harmonics linked to the crystal symmetry.**

The spatial profiles of high harmonics are crucial information for characterising the mixing of spin states through the SOI of light. Figure 3a displays the RCP and LCP components of the spatial profiles of the second, third, fourth, and fifth-order harmonics obtained under the tight focus condition. In contrast to the circularly symmetrical incident beam, a variety of structured light appeared. In the LCP components of the second and fifth-order harmonics, and RCP component of the fourth-order harmonics, bright spots appeared at the centre of the beam, signifying the presence of the $l$=0 mode. These polarization components are present even in the absence of SOI[46]. Although the other components in Fig. 3a should be forbidden in the absence of SOI, donut-shaped patterns were clearly observed in our experiments. These observations imply that non-zero OAM modes were generated. We also observed characteristic structures with six nodes in the azimuthal direction. These observations are much different from those acquired under the loose focus condition, where only circular profiles were observed (see Fig. S1 in the Supplementary Information). To understand the observed spatial patterns, we reproduced the structures as shown in Fig. 3b by fittings consisting of sums of Laguerre-Gaussian modes. These successful fits suggest that the six-fold structures result from interference between similar amounts of different OAM modes with indexes separated by six. We confirmed the inversion of the spiral structures by inverting the polarization of the fundamental beam (see Fig. S2). This symmetric behaviour corresponds to reversing the sign of the relative phase difference between the radial modes.

The observed symmetric light structures show an unprecedented link to the crystal's structure. When we rotated the GaSe crystal around the optical axis, the spatial profiles of the harmonics rotated accordingly, as represented by fourth-order RCP harmonic in Fig. 4. This rotation corresponds to a shift in the relative phases between the multiplexed OAM modes. Note that this link is unique to phenomena in the combination of nonlinear optics and SOI. In linear optics, the details of the crystal structure do not manifest themselves in the optical response, with only the refractive index influencing its behaviour. In nonlinear optics, however, the polarization of light is sensitive to the symmetry of the crystal. Furthermore, through SOI, the polarization correlates with macroscopic spatial light structures. Therefore, the combination of nonlinear optics and SOI allows us to control macroscopic optical responses through microscopic light-matter interactions.

**Identification of orbital angular momenta for comparison with conservation rule.**

To identify the OAM forming the structured harmonics, we disentangled it with a spatial light modulator (SLM), as illustrated in Fig. S5. By applying additional phase factors $\exp(i\Delta l\phi)$ to the collimated harmonics depending on the azimuthal phase $\phi$, the spin-orbit states transformed as $(s_m, l_m) \rightarrow (s_m, l_m + \Delta l)$. Subsequently, the spatial image of the reflected high harmonics was Fourier-transformed by a lens, and the resulting image was detected by a camera. The OAM contained in the harmonics were identified by observing the central spot that became bright when $\Delta l$ equalled $-l_m$ [44].

Figures 5a displays measured images of harmonics of all orders by applying $\Delta l$s from 2 to -10 to both circular polarization components. Here, the spatial profiles varied in accordance with the phase variations of the SLM. The images enclosed in the red squares are spatial patterns whose intensity distribution concentrates at the centre. These images directly correspond with the OAM components $l_m = -\Delta l$ forming the harmonics. For example, they indicate that the fifth-order RCP component is a superposition of three OAM states, $l = -2, 4, 10$. Notably, the positions of the red squares are limited to even-number values of $l_m$ and show clear patterns with respect to OAM $l_m$ and the harmonic order $m$, dependent on the polarization. This result indicates the presence of angular momentum selection rules.

Further clues to the interpretation of the results were obtained by examining the radial dependence of the angle-averaged harmonic intensity. In Fig. 5b, the fourth-order RCP components depict clear increases in the radius of the intensity distribution as $\Delta l$ deviates from 0 and 6. This observation accords with the fact that OAM modes with larger $l_m$ exhibit a ring-shaped intensity distribution with a larger radius in the focal plane. This finding supports the description of the obtained harmonics as a sum of OAM components with $l = 0$ and 6. Similar results support the presence of $l = 2$ and 8 for the fourth-order LCP components (Fig. S3 shows results for other orders).

Figure 5c summarises the observed angular momentum states obtained by integrating the harmonic intensity around the centre of the images in Fig. 5a. Interestingly, all of the experimental data points show clear peaks that lie at the conditions predicted by the total angular momentum conservation rule (4) with $n = 3$ and the conditions restricted by the equation (5).

**Discussion**

The measured OAM demonstrates that the total angular momentum is conserved even in the presence of complex nonlinear processes and propagation effects in the crystal. The cascading processes involving nonlinear conversion and SOI (Fig. 6a) provide fundamental insights into the dynamics of angular momentum and frequency conversion of light. Here, we denote the state of the spin and orbital angular momentum in the $m$-th order harmonics as $(m; s, l)$ to illustrate energy and angular momentum conservations. For example, when light propagates in the crystal, part of a fundamental light $(1; 1, 0)$ is converted into the $(1; -1, 2)$ component due to the SOI[41,42]. The observed harmonics are expected to come primarily from these two photons. In conventional nonlinear optics, the OAM of the harmonics arises from the sum of the OAMs of the photons involved in generating the harmonics. This explains why the majority of OAM components in Fig. 5 have positive signs. For example, the presence of the $(5; 1,10)$ component can be attributed to the fifth multiple of the $(1; -1,2)$ component[46,50]. However, generating harmonic components with negative OAM is more complicated. For example, generating $(3; -1, -2)$ requires three steps, involving second harmonic generation of $(2; -1,0)$ from $(1; 1,0)$, generation of $(2; 1, -2)$ from $(2; -1,0)$ via the SOI, and sum frequency generation of $(3; -1, -2)$ from the $(2; 1, -2)$ and $(1; 1,0)$. Thus, our observation of negative OAM components reveals that the cascade process makes a crucial contribution to HHG in bulk crystals even for above-bandgap nonperturbative harmonics. The grey shaded area in Fig. 5c shows the OAM components that can be generated by the above-mentioned cascade processes. Most of the observed components lie around the central part of the shaded area, since there are more cascade conversion paths to create OAM components around the central part than those around edge of the shaded area. Note that equations (2) and (3) were derived by considering only the interaction between the fundamental beam and the crystal. If symmetry is maintained at a given z-section in the crystal, then the nonlinear polarization at that section will have the same symmetry. Consequently, at a section $z + \Delta z$, the total electric field can have different spatio-temporal profiles from those at $z$, while still maintaining the same DS. This property ensures the robustness of DS even in the presence of complex cascading processes.

In summary, we observed high harmonics with a spatial structure linked to the crystal symmetry of solids by incorporating the SOI of light. We showed that the modified total angular momentum conservation rule, reflecting discrete crystal symmetry, provides essential insights into the spin-dependent OAM control in general nonlinear processes, including cascade and extreme nonlinear phenomena. Moreover, we demonstrated that multiscale dynamical symmetry effectively works as the combined symmetry of solid and light in these phenomena. Our results pave the way for solid-based engineering of structured light pulses and exploration of their topological properties in the extreme ultraviolet region[5,52]. In addition, the dynamic modulation of Floquet states by an intense pulsed laser field may offer a method for ultrafast temporal shaping of vectorial structures[30,31,53]. Furthermore, synchronizing the symmetry of crystals with metasurfaces, photonic crystals, and optoelectronic devices may present ways of expanding the functionalities of solids in nonlinear photonics[19,27,32,33,54].

**Acknowledgements**

This work was supported by a Grant-in-Aid for Scientific Research (S) (Grant No. JP20H05670).

**Author Contributions**

K. N. conceived the project and carried out the experiment with assistance from T.O. Y. S. assisted with the theoretical calculation and discussed the results. H. S. and K. O. supervised the project. All authors contributed to the scientific discussion and writing of the manuscript.

**Methods**

**Experimental setup**

The infrared HHG driver was generated by a BBO-based optical parametric amplifier (TOPAS-prime, Light Conversion) pumped by a Ti: sapphire amplifier (centre wavelength, 800 nm; pulse energy, 3 mJ; pulse duration, 20 fs; repetition rate, 3 kHz). The centre wavelength of the driver source was around 2.4 μm and the pulse duration was 80 fs as estimated from the Fourier limit at full width of half maximum (FWHM). To eliminate the undesired beam from the OPA, the spectra shorter than a wavelength of 1650 nm was blocked by a longpass filter. The nearly collimated infrared driver source was spatially filtered with a 2.5 mm aperture iris to shape the beam profile as close to a Gaussian profile as possible. The power of the fundamental beam was controlled by wire-grid polarizers, and the beam was converted to a circularly polarized one with an achromatic quarter-wave plate (SAQWP05M-1700, Thorlabs). To achieve an external divergence angle of 402 mrad, which satisfies the conditions for significant oblique incidence to induce SOI of light in a 2-mm-thick GaSe crystal (Eksma optics), the fundamental beam with a diameter of 3 mm (FWHM) was focused by using an aspherical lens with a focal length of 6 mm. The resultant high harmonics from the GaSe were refocused by an objective lens (x20, Nikon) onto a spectrometer (QE-pro, Ocean Insight) or a color complementary metal-oxide semiconductor (CMOS) camera (CS165CU/M, Thorlabs) to obtain harmonic spectra and spatial profiles. The focal position of the beam relative to the crystal was optimized to maximize the intensity of the sixth harmonic. The polarization of the high harmonics was analyzed by an achromatic quarter-wave plate (SAQWP05M-700, Thorlabs) and a wire-grid polarizer (WP25M-UB, Thorlabs). In order to acquire the spatial images of the high harmonics, pairs of shortpass and longpass filters were inserted in front of the CMOS camera to pick up each order of harmonics. Since the peak wavelength of the second harmonics lies on the edge of the detectable spectral region of Si sensors, the short wavelength tale of the second harmonics around 1000 nm was detected by the camera. The focused beam spot size was estimated to be 4 μm from the spot size of the LCP component of the second harmonic generation. The estimation assumes that the intensity of the second harmonics is proportional to the square of the intensity of the fundamental beam and equivalent amounts of RCP and LCP components are present at the focal plane. The intensity at the focus point was estimated by using this beam spot size. In the loose-focus setup designed for comparatively showing the effect of SOI, an incident beam with a Gaussian divergence angle of 12.7 mrad was selected using a lens with a focal length of 200 mm. This condition results in a relatively large beam waist in the crystal and lower beam intensity. We therefore used a higher incident pulse energy of 3.7 uJ to generate the high harmonics. For the collimation of the high harmonics in this setup, an achromatic lens with a focal length of 30 mm was used instead of the objective lens.

The OAM states of the harmonics were disentangled by using a spatial light modulator (SLM), (SLM-200, Santec). The optical setup for using the SLM is shown in Fig. S5. The harmonics were collimated by the objective lens and were reflected by the SLM to apply an additional azimuthal phase to the harmonics. The reflected light beam was focused by a lens onto the CMOS camera to apply a Fourier transformation of the spatial patterns of the beam. To control the wavefront of the second, third, fourth, and fifth harmonics, corresponding phase patterns at wavelengths of 1000 nm, 800 nm, 600 nm, and 480 nm were displayed on the SLM.

**Analysis of color images**

All color images of the harmonics are RGB color data. For visibility, the spatial images of the harmonics were normalized by their maximum values after subtraction of the background signal and then processed by a gamma correction with a Γ value of 2.2. The white balance of the RGB data was determined by multiplying the raw data obtained by the color CMOS camera, which had the same gain for each color sensor, by coefficients of 2.7, 1, and 3.15.

The fitting of the 2D images in Fig. 3 was performed by considering Laguerre-Gaussian modes. The fitting functions were

$$S(\rho, \phi) = \left| \sum_{p,l} C_{p,l} U_{p,l}(\rho) e^{il\phi} \right|^2$$

with

$$U_{p,l}(\rho) = \sqrt{\frac{2p!}{\pi(|l|+p)!}} \frac{1}{w_0} \left(\frac{\rho\sqrt{2}}{w_0}\right)^{|l|} L_p^{|l|}\left(\frac{2\rho^2}{w_0^2}\right) \exp(-\frac{\rho^2}{w_0^2}),$$

where $L_p^l$ denotes the associated Laguerre polynomial. The indices $p$ and $l$ denote the radial mode and OAM mode, respectively. Polar coordinates around the centre pixel in the images are defined by the radial position ρ and azimuthal angle φ. The fitting parameters were the complex coefficients $C_{p,l}$, beam waist $w_0$, and centre pixel in the images. Two OAM modes and three radial modes were considered. The OAM modes were determined so as to match the two shown in Fig. 5. For the second-order LCP and third-order RCP components, only single OAM modes were considered for the fitting, as the interference structures between different OAM modes were absent in these observed images. The fitting was performed on 2D data obtained by averaging the three values for RGB colors. The relative intensities between the RGB colors were determined to be those that fit the experimental results. All parameters used to reconstruct Fig. 3 are shown in Fig. S2.

**Calculation for spin-orbit mixing of fundamental beam**

Our estimation of the power exchange between (RCP, $l = 0$) and (LCP, $l = -2$) relies on the Ciattoni-Cincotti-Palma scheme[41] for a paraxial beam propagating along the optical axis of a uniaxial medium. In the related literature[41,42], the mixing of spin- and orbital-angular momenta of light is assumed to happen at the beam waist of a Gaussian beam. On the contrary, the mixing starts from a plane off the beam waist in our experimental situation. We customized the derived formula to be applicable to our experimental situation as described below.

The incident electric field in the vacuum is assumed to be a Gaussian beam $\boldsymbol{E}^{\text{vac}}(r, \phi, z, t) = \Re(\boldsymbol{e}_+ E_{\text{Gb}}(r, \phi, z) e^{-i\omega t})$, $\boldsymbol{e}_+ = (x + iy)/\sqrt{2}$ propagating along the z-axis, where $(r, \phi)$ is the radial coordinate for the x-

$y$ plane. The Gaussian envelope focused on $z_f$ with waist $w_0$ is expressed as $E_{\text{Gb}}(r,\phi,z) = E_0[w_0/w(z-z_f)]e^{-r^2/w^2(z-z_f)}e^{ik_0(z-z_f)}e^{ik_0r^2/2R(z-z_f)}e^{-i\psi(z-z_f)}$, where $w(z) = w_0\sqrt{1+(z/z_R)^2}$, $z_R = \pi w_0^2/\lambda$, $1/R(z) = z/(z^2+z_R^2)$, $\psi(z) = \arctan(z/z_R)$. We put the sample at $0 \leq z < 2$ mm. The complex field envelope is

$$E_{\text{Gb}}(r,\phi,z=0) = \frac{E_0' w_0}{w(-z_f)} e^{-r^2/2s^2(-z_f)}, \quad s^2(z) = \left[\frac{2}{w^2(z)} - \frac{ik_0}{R(z)}\right]^{-1} = \frac{w_0^2}{2}\left(1 + i\frac{z}{z_R}\right)$$

Where the irrelevant phase not depending on $r$ is absorbed in the phase of $E_0'$. The complex waist squared function $s^2(z)$ becomes real-valued only at the focal point. According to Eq. (1) and (2) in Ref. 41, the field over the whole medium is determined by the Fourier transformation on the $x$-$y$ plane at $z=0$. The Fourier transformation of our Gaussian envelop is $\tilde{E}_{\text{Gb}}(k,z=0) = [E_0' w_0 s^2(-z_f)/2\pi w(-z_f)]e^{-k^2 s^2(-z_f)/2}$. We obtain the absolute square as

$$\left|\tilde{E}_{\text{Gb}}(k,z=0)\right|^2 = \frac{|E_0'|^2 w_0^2 |s^2(-z_f)|^2}{(2\pi)^2 w^2(-z_f)} e^{-k^2 \Re[s^2(-z_f)]} = \frac{|E_0'|^2 w_0^2}{16\pi^2} e^{-k^2 w_0^2/2}.$$

One should note that the absolute square does not depend on $z_f$ any more. The powers of RCP(+) and LCP(-) components, Eq. (22) and (B4) in Ref. 41, is given as

$$W_\pm(z) = \frac{1}{2} W_{\text{tot}} \pm 4\pi^3 \int_0^\infty dk\, k \cos\left(\frac{z\Delta}{2k_0 n_o}k^2\right) |\tilde{E}_{\text{Gb}}(k,z=0)|^2, \quad W_{\text{tot}} = 8\pi^3 \int_0^\infty dk\, k |\tilde{E}_{\text{Gb}}(k,z=0)|^2,$$

where $\Delta = n_e^2/n_o^2 - 1$. We obtain the powers of the RCP and LCP components as functions of $z$ as

$$\frac{W_\pm(z)}{W_{\text{tot}}} = 1 \pm \frac{1}{1+(z/L)^2}, \quad W_{\text{tot}} = \frac{\pi|E_0'|^2 w_0^2}{2}, \quad L = \frac{k_0 n_o w_0^2}{\Delta}.$$

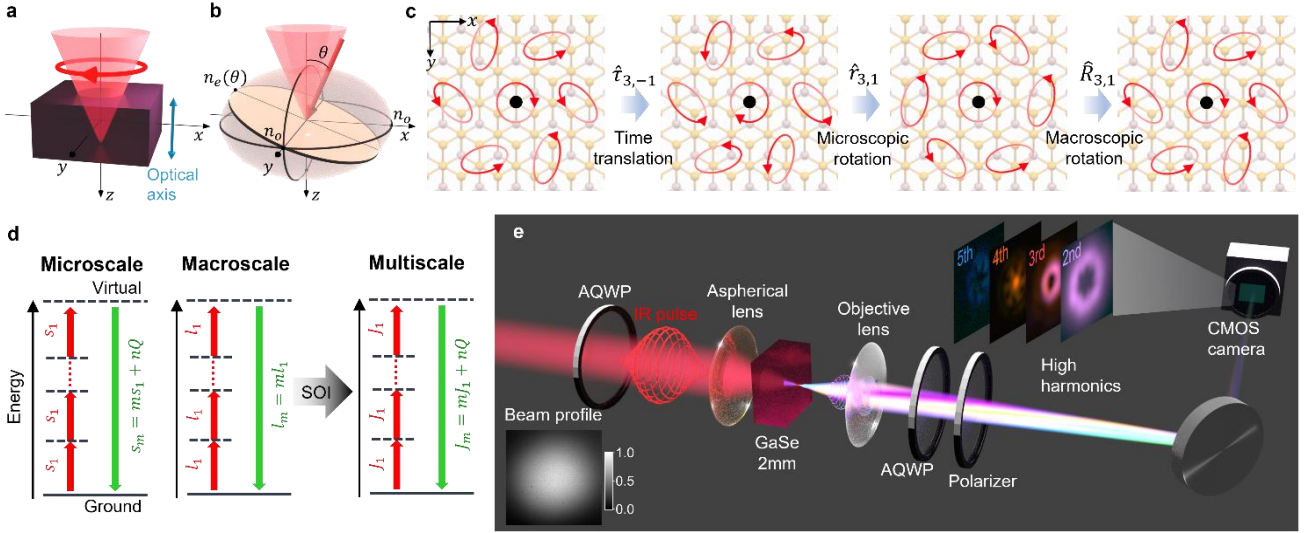

**Fig. 1. Multiscale dynamical symmetry created by spin-orbit interaction of light in uniaxial solids. (a)** Tight-focus configuration of circularly polarized light in a uniaxial crystal. **(b)** Index ellipsoid of negative uniaxial crystal. The refractive indices on the extraordinary and ordinary axes are denoted by $n_e$ and $n_o$, respectively. Wavevector components inclined by an angle $\theta$ relative to the optical axis experience axisymmetric birefringence with refractive indices of $n_e(\theta)$ and $n_o$. **(c)** Multiscale dynamical symmetry operations on spatial distribution of the polarization state of the driving light field and GaSe crystal in a plane parallel to the *x-y* plane inside the crystal. The fundamental beam is assumed to have right-circular polarization ($s_1 = 1$). The seven ellipsoids with arrows represent the polarization and phase of the laser electric field at each spatial point. Subsequent operations of the time translation $\hat{\tau}$, the microscopic operation $\hat{r}$, and the macroscopic rotation $\hat{R}$ make the system consisting of light and solid identical to the original. **(d)** Photon diagram for angular momentum conservation in high harmonics. Spin ($s_m$) or orbital angular momentum ($l_m$) is conserved when micro- or macroscopic symmetry is present. Spin-orbit interaction (SOI) of light entangles the angular momenta, leaving only the total angular momentum $J_m$ conserved. Dashed lines represent virtual states, and the solid line represents the ground state in the electronic transitions. **(e)** Experimental configuration for the imaging of spin-angular-momentum-resolved high harmonics from GaSe crystal. The lower-left inset is the beam profile of the fundamental beam measured before it is focused with an aspherical lens. Normalized intensity is shown on a linear color scale. AQWP: achromatic quarter wave plate.

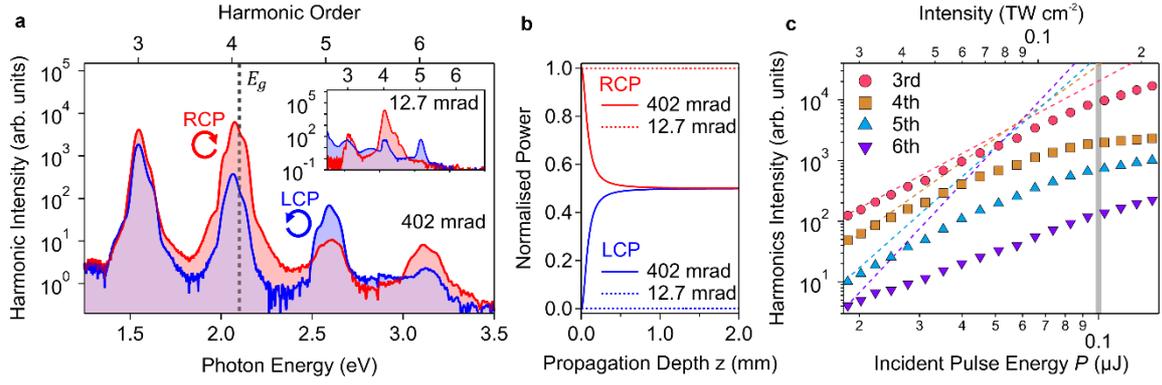

**Fig. 2. High harmonics in extreme nonlinear regime generated by tightly focused paraxial Gaussian beam.** (a) Circular-polarization-resolved harmonic spectra from GaSe crystal up to the sixth order. External divergence angle of incidence was chosen to be 402 mrad by using an aspherical lens with a focal length of 6 mm. Right-circularly polarized (RCP) and left-circularly polarized (LCP) components are shown in red and blue, respectively. The dashed line represents the bandgap energy $E_g$ of the GaSe crystal; an incident pulse energy of 0.1 uJ was used. Inset: spectra corresponding to loosely focused driving beam with an external beam divergence angle of 12.7 mrad. A lens with a focal length of 200 mm and an incident pulse energy of 3.7 uJ was used for the loose focus. (b) Calculated power transfer between RCP and LCP components of the fundamental wave in the GaSe crystal due to spin-orbit interaction of light. The red and blue lines are the powers of the RCP and LCP components, respectively, in a slice at a depth of z from the front surface of the GaSe crystal. These powers are normalised by the total power of the fundamental wave. Solid and dashed lines represent calculated results for Gaussian external divergence angle of 402 mrad and 12.7 mrad. (c) Dependence of the harmonic intensity on the incident pulse energy $P$. The grey line represents the incident power used in the experiments. Input pulse energy of 0.1 μJ corresponds to the intensity of 0.15 TW cm$^{-2}$ at the focal plane in vacuum. Each colored dashed line is a guide for the eye showing the power law, where the $m$-th harmonic intensity is proportional to $P^m$.

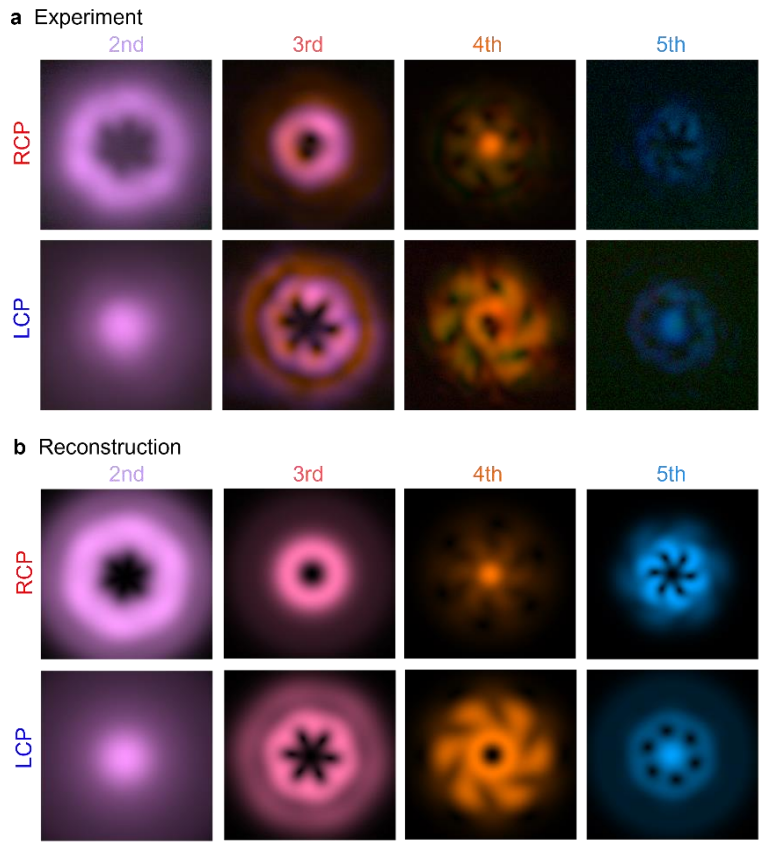

**Fig. 3. Spatial imaging of vectorially structured high harmonics.** (**a**) Spatial profiles of right and left circularly polarized components of high harmonics. (**b**) Reconstructed spatial profiles made from fittings with two orbital angular momentum and three radial modes of the Laguerre-Gaussian series.

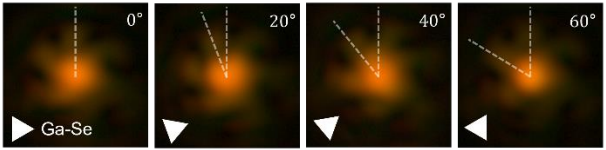

**Fig. 4. Macro- and microscopic structures of light and crystal linked through spin-orbit interaction**. Spatial profiles of right-circularly polarized (RCP) component of the fourth harmonic at different crystal orientations. Orientation of Ga-Se bonding in the crystal is represented by the sides of white triangles. White dashed lines are guides for the eye that represent the azimuthal phase of spiral structures.

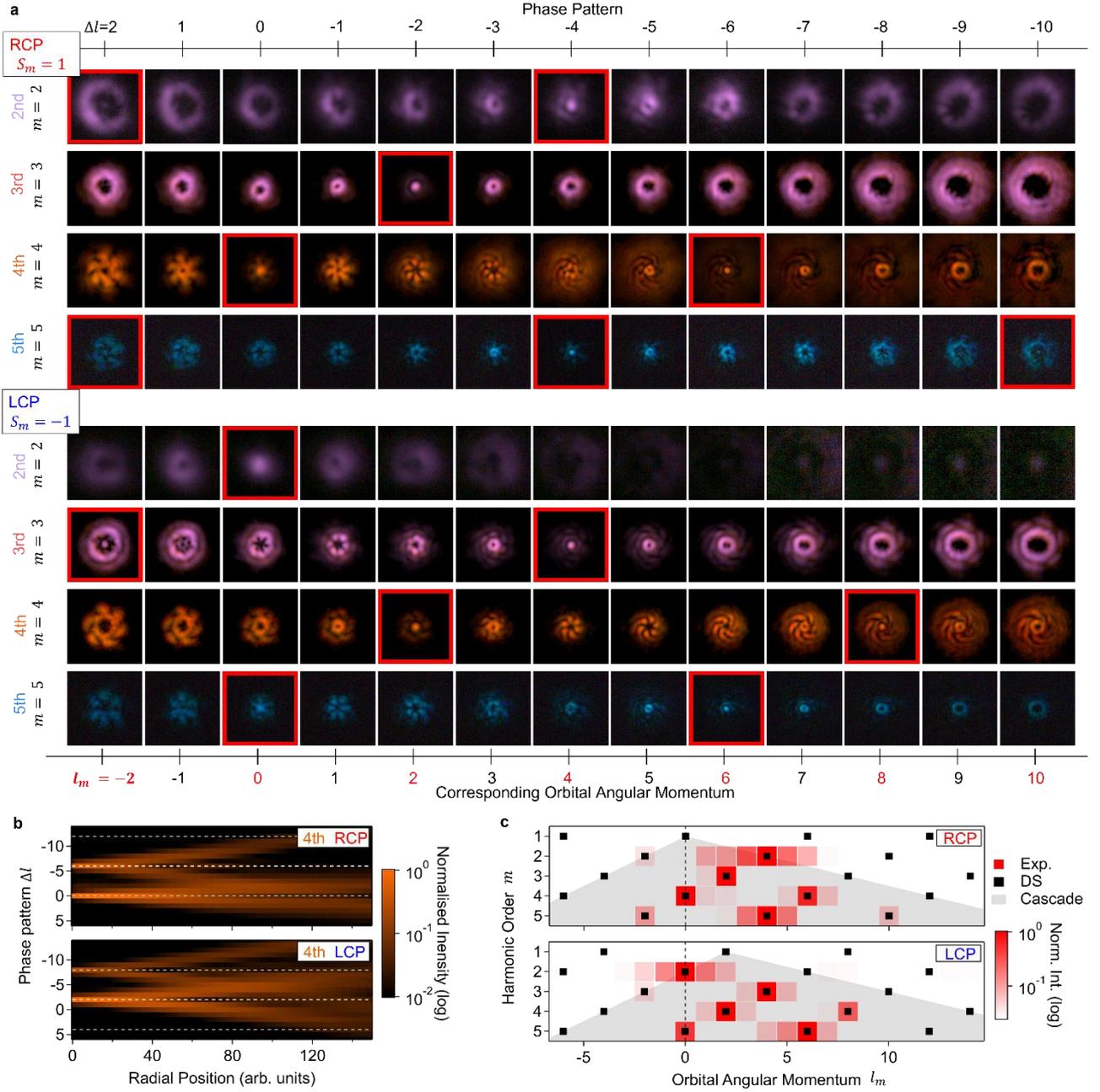

**Fig. 5. Spin-orbit tomography for high harmonics and selection rule. (a)** Imaging of the phase-modulated harmonics reflected from spatial light modulator (SLM) through Fourier transformation by a lens. The red squares represent experimental data that show bright spots at the beam centre. Bottom axis shows orbital angular momentum corresponding to the phase patterns displayed on the SLM, i.e., $l_m = -\Delta l$. **(b)** Azimuthal-angle averaged plots for **(a)** as a function of radial positions for the fourth-order harmonics (upper: right circular polarization (RCP) component, lower: left circular polarization (LCP) component). To determine the original points of the polar coordinates, the centre positions of the respective images are chosen for the respective harmonic order and polarization components. Intensity normalized for each polarization is shown on a log color scale. White dashed lines represent OAM components allowed by the total angular momentum conservation rule derived from the multiscale dynamical symmetry (DS) in equation (4). **(c)** Table of spin and orbital angular momenta of light in high harmonics. The red-square experimental data points are calculated by integrating the harmonic signal in $10 \times 10$ pixels on the centre spots of the original $740 \times 740$ pixel images in **(a)** (upper: RCP, lower: LCP). Intensity normalized for each harmonic order and polarization component is shown on a log color scale. Black squares represent the allowed states by the total angular momentum conservation rule derived from multiscale dynamical symmetry. The grey area indicates the region where components allowed by the cascaded process of harmonic generation (HG) and spin-orbit interaction (SOI) of light are present.

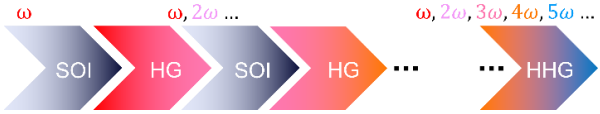

**Fig. 6. High harmonic spin-orbit angular momentum cascade.** Diagram of cascade process of spin-orbit interaction (SOI) of light and harmonic generation (HG) in a uniaxial crystal. High harmonic generation (HHG) process at the back surface of the crystal is driven by the total electric field generated by the cascade process. SOI affects both fundamental and harmonic waves during the propagation.

# Supplementary Information for
# Crystal-symmetry control of vectorially structured high harmonics

Kohei Nagai, Takuya Okamoto, Yasushi Shinohara, Haruki Sanada, and Katsuya Oguri

NTT Basic Research Laboratories, NTT Corporation, 3-1, Morinosato-Wakamiya, Atsugi-shi, Kanagawa 243-0198, Japan

e-mail: kouhei.nagai@ntt.com

## I. Derivation of total angular momentum conservation rule.

When electron systems in solids interacting with light has a multiscale DS described by an operator $\hat{G}$, the symmetry of the system is reflected in the electric field $\vec{E}$ of emitted harmonics as follows:

$$\hat{G}\vec{E}(\vec{R},t) = \vec{E}(\vec{R},t), \tag{S1}$$

where $\vec{R}$ denotes a macroscopic position vector. To consider the operator $\hat{G}_1$ in equations (2), we take a cylindrical coordinate for the system. By considering the temporal periodicity of the electric field, we can provide an ansatz for the electric field in an x-y plane using Fourier expansion:

$$\vec{E}(\vec{R},t) = \sum_{m,s_m,l_m} A_{m,s_m,l_m}(\rho,z) e^{im\omega t + il_m\phi} \begin{pmatrix} 1 \\ -is_m \end{pmatrix}, \quad \vec{R} = \rho(\cos\phi\,\vec{x} + \sin\phi\,\vec{y}) + z\vec{z}, \tag{S2}$$

where $A_{m,s_m,l_m}(\rho,z)$ represents the amplitude of each spin and orbital angular momentum mode of $m$-th order harmonic field. By applying the DS operator $\hat{G}_1$ in equations (2) to (S2), we get

$$\sum_{m,s_m,l_m} A_{m,s_m,l_m}(\rho,z) e^{im\omega t + il_m\phi - \frac{i2\pi m(l_1+s_1)}{n} + \frac{i2\pi l_m}{n} + \frac{i2\pi s_m}{n}} \begin{pmatrix} 1 \\ -is_m \end{pmatrix}$$

$$= \sum_{m,s_m,l_m} A_{m,s_m,l_m}(\rho,z) e^{im\omega t + il_m\phi} \begin{pmatrix} 1 \\ -is_m \end{pmatrix}. \tag{S3}$$

Since this equation holds for each Fourier component, we obtain equation (4). Moreover, applying $\hat{G}_2$ in equation (3) to equation (S2), we obtain the restriction of $l_m$ as shown in equation (5)

## II. Comparison between the spatial profiles of high harmonics in the tight- and loose-focus setup.

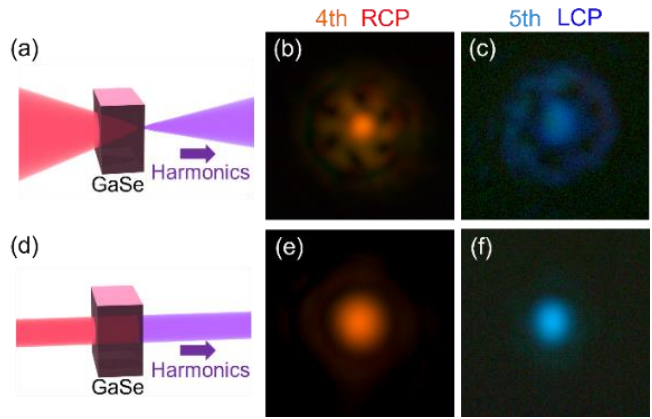

**Figure S1: Comparison between spatial profiles of high harmonics in tight- and loose-focus setup.** (a) Experimental configuration of the tight focus. Spatial profiles of (b) fourth order RCP and (c) fifth order LCP harmonics. (d-f) same as (a-c) for the loose focus.

## III. Fitting results for Fig. 3.

| 2nd RCP | Amplitude | Phase | Amplitude | Phase |
|---|---|---|---|---|
| | l=4 | | l=-2 | |
| p=0 | 1 | 0 | 0.01199137 | -2.4010952 |
| p=1 | 0.1097001 | -2.5714371 | 0.038106076 | 2.0811727 |
| p=2 | 0.40131569 | 1.2292758 | 0.010120554 | 3.1189749 |

| 2nd LCP | Amplitude | Phase |
|---|---|---|
| | l=0 | |
| p=0 | 1 | 0 |
| p=1 | 0.29691884 | -2.5068967 |
| p=2 | 0.41599074 | -0.81192648 |

| 3rd RCP | Amplitude | Phase |
|---|---|---|
| | l=2 | |
| p=0 | 1 | 0 |
| p=1 | 0.96661735 | 0.52935755 |
| p=2 | 0.48406097 | 1.688997 |

| 3rd LCP | Amplitude | Phase | Amplitude | Phase |
|---|---|---|---|---|
| | l=4 | | l=-2 | |
| p=0 | 0.91413605 | 0 | 0.14567097 | -1.2613226 |
| p=1 | 1 | -1.1603539 | 0.16433184 | -1.3515117 |
| p=2 | 0.1840733 | 2.6860237 | 0.11149683 | -1.9677181 |

| 4th RCP | Amplitude | Phase | Amplitude | Phase |
|---|---|---|---|---|
| | l=6 | | l=0 | |
| p=0 | 0.23151776 | 0 | 1 | -0.42846531 |
| p=1 | 0.18283522 | -1.2580431 | 0.32996419 | -0.92983949 |
| p=2 | 0.13481876 | 1.2407612 | 0.39002851 | -1.2137977 |

| 4th LCP | Amplitude | Phase | Amplitude | Phase |
|---|---|---|---|---|
| | l=8 | | l=2 | |
| p=0 | 0.2508342 | 0 | 0.54007387 | 0.68371737 |
| p=1 | 0.14972655 | 0.0016975929 | 1 | -0.45262063 |
| p=2 | 0.094694287 | -2.4344046 | 0.51291043 | -2.2247744 |

| 5th RCP | Amplitude | Phase | Amplitude | Phase |
|---|---|---|---|---|
| | l=4 | | l=-2 | |
| p=0 | 1 | 0 | 0.28339803 | -0.42846531 |
| p=1 | 0.29128331 | -1.2580431 | 0.18998504 | -0.92983949 |
| p=2 | 0.17217995 | 1.2407612 | 0.22389999 | -1.2137977 |

| 5th LCP | Amplitude | Phase | Amplitude | Phase |
|---|---|---|---|---|
| | l=6 | | l=0 | |
| p=0 | 1 | 0 | 0.96352166 | -2.5870135 |
| p=1 | 0.35661048 | -1.0686089 | 0.32130644 | -2.8193302 |
| p=2 | 0.61791688 | -0.48705125 | 0.03169819 | 2.4371681 |

**Figure S2: Fitting results for Fig. 3.** Each table shows normalised amplitude and phase (rad) of each Laguerre-Gaussian mode to reproduce the experimental results in Fig. 3a. Three radial modes indicated by $p$ are considered to reproduce approximate spatial structures of harmonics. The values of OAM $l$ is determined to match the dominant modes present in the results shown in Fig. 5. Second-order left-circularly polarized (LCP) and third-order right-circularly polarized (LCP) components are fitted with single OAM modes.

## IV. Inverted spatial structures of harmonics in time-reversal condition.

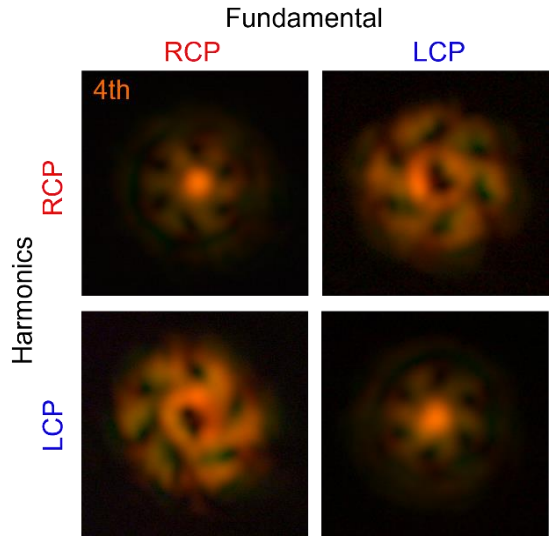

**Figure S3: Inverted spatial structures of harmonics in time-reversal condition.** Comparison of the spatial profiles of the fourth order harmonics between the case with right (RCP) and left circularly polarized (LCP) fundamental beam (comparison between left two panels and right two panels). The polarization of the fundamental beam was controlled by inverting an achromatic quarter wave plate for the fundamental beam. Upper and lower two panels show RCP and LCP components of the fourth order harmonic, respectively.

## V. Impact of Gouy phase around focal plane.

Relative phases between different OAM modes are varied around Rayleigh range due to the effect of Gouy phase. The Gouy phase for Laguerre-Gaussian modes are written by

$$\chi_{l,p}(z) = (2p + |l| + 1)\arctan(z/z_R)],$$

where $p$ and $l$ are radial and azimuthal index for the Laguerre Gaussian modes, respectively and $z_R$ is the Rayleigh range. Due to the different z-dependence of the Gouy phase between different OAM modes, the phase of the interference patterns undergoes rotation by changing z around the focusing point. In fact, in experimental observations, the azimuthal node structures show rotation with changing the focus point by the z-position of the objective lens. This observation supports that the obtained high harmonics are indeed composed of a superposition of OAM modes.

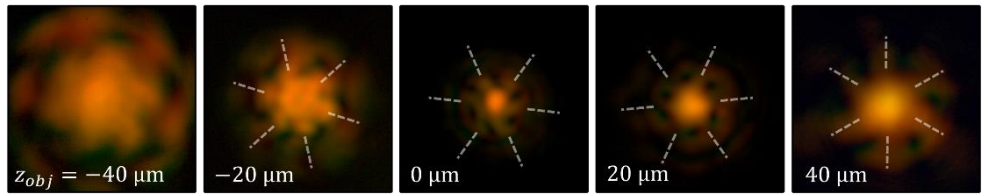

**Figure S4: Spatial profiles of the fourth order harmonics with respect to the position of the objective lens.** Measured spatial profiles for the right circularly polarized component corresponding to the z-positions of the objective lens at $z_{obj} = -40, -20, 0, 20, 40$ μm. Dashed lines are eye guides to represent the azimuthal phase of spiral structures. The absolute position of $z_{obj} = 0$ was determined so that the spatial structures of the harmonics were minimised on the camera.

## VI. Schematics of optical setup for spin-orbit tomography.

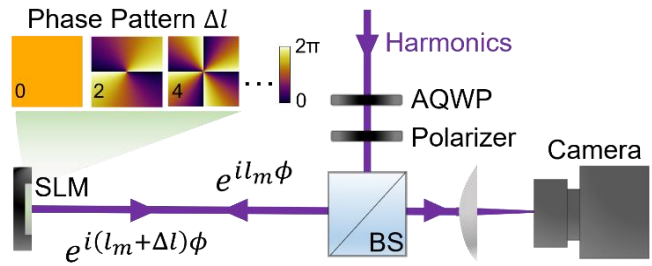

**Figure S5: Optical setup for spin-orbit tomography** QWP: achromatic quarter wave plate, SLM: spatial light modulator, BS: non-polarizing beam splitter.

# VII. Disentangling orbital angular momentum components that form second, third and fifth harmonics.

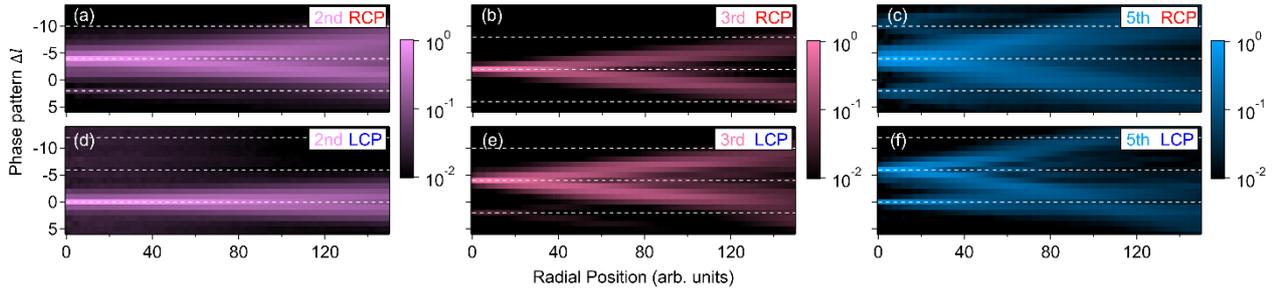

**Figure S6: Azimuthal-angle averaged plots for normalized harmonic intensity as a function of radial positions in the images in Fig. 5a.** Results for (a,b,c) right-circularly polarised and (d,e,f) left-circularly polarized components of (a,d) second, (b,e) third, (c,f) fifth order harmonics. Normalized intensity is shown on log color scales. A weak signal insensitive to the SLM phase pattern is observed (also shown in Fig. 5a). This signal may arise from multiple internal reflections within the beam splitter in Fig. S5.